# Major transitions in evolution linked to thermal gradients above hydrothermal vents

Anthonie W.J. Muller

**The emergence of the main divisions of today's life: (1) unicellular prokaryotes, (2) unicellular eukaryotes, (3) multicellular eukaryotes, and (4) metazoans, are examples of the—still unexplained[1]—major transitions in evolution[2]. Regarding the origin of life, I have proposed that primordial life functioned as heat engine ('thermosynthesis') while thermally cycled in convecting volcanic hot springs[3]. Here I argue for a role of thermal gradients above submarine hydrothermal vents (SHV)[4] in several major transitions. The last decade has witnessed the emergence of phononics[5], a novel discipline in physics based on controlled heat transport in thermal gradients. It builds thermal analogs to electronic devices: the thermal diode, the thermal transistor, the thermal switch, the thermal amplifier, the thermal memory—the thermal computer has been proposed. Encouraged by (1) the many similarities between microtubules (MT) and carbon nanotubes[6], which have a very high thermal conductivity[7], and (2) the recent discovery of a silk protein which also has a very high thermal conductivity[8], I combine and extend the mentioned ideas, and propose the *general conjecture* that *several major transitions of evolution were effected by thermal processes*, with four additional partial conjectures: (1) The first organisms used heat engines during thermosynthesis in convection cells; (2) The first eukaryotic cells used MT during thermosynthesis in the thermal gradient above SHV; (3) The first metazoans used transport of water or in water during thermosynthesis above SHV under an ice-covered ocean during the Gaskiers Snowball Earth; and (4) The first mammalian brain used a thermal machinery based on thermal gradients in or across the cortex. When experimentally proven these conjectures, which are testable by the methods of synthetic biology, would significantly enhance our understanding of life.**

---

Swammerdam Institute for Life Sciences, University of Amsterdam, 1098XH Amsterdam, The Netherlands

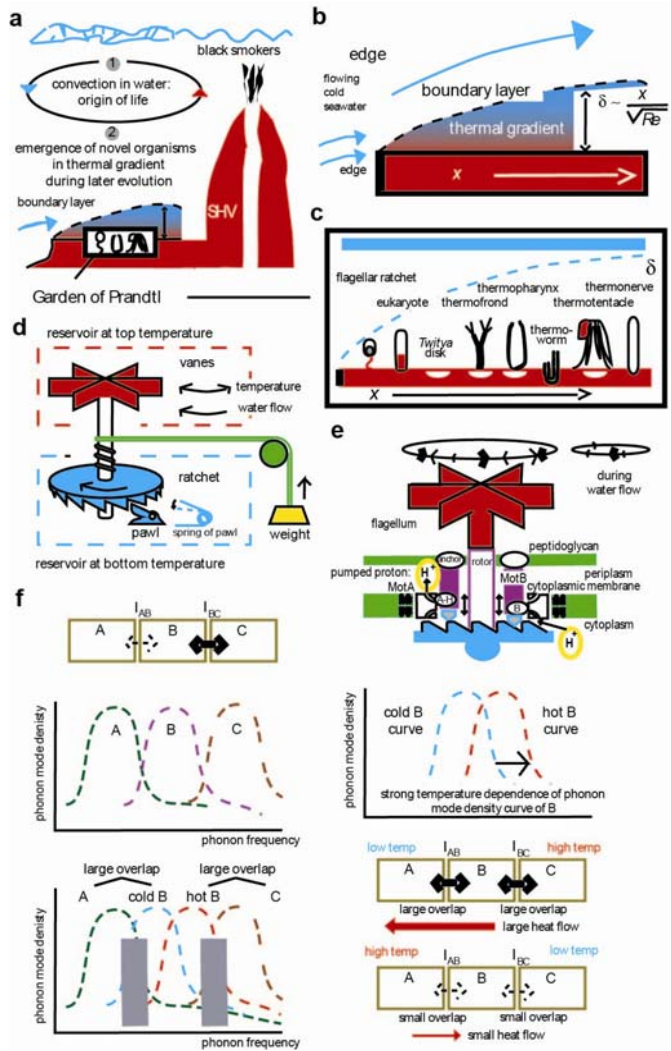

**Figure 1. Overview of invoked thermal mechanisms. a,** Convection cell and thermal gradient near SHV. Where water flows over raises above the ocean floor, the boundary or Prandtl layer with laminar flow spans the thermal gradient. **b,** The boundary layer thickness δ increases with distance $x$ to the edge by $\delta \sim x/\sqrt{Re}$, with $Re$ is the Reynolds number of the flow[9]. **c,** In the Garden of Prandtl the smallest organisms emerged first, at small $x$ and δ, and larger organisms later. **d,** Feynman's ratchet[10] consists of a ratchet-and-pawl connected to vanes, both subject to a Brownian fluctuating force. When the vanes are colder than the ratchet-and-pawl, the weight drops and forces the vanes to rotate while the hot pawl repeatedly jumps the ratchet, but when warmer, the weight is raised, and the system functions as heat engine. When the axis is forced to rotate in the indicated direction by a torque due to water flow on the flagellar 'propeller', the weight raises as well: an external water flow mimics a high temperature. **e,** Previously proposed proton-pumping flagellar ratchet[9], analog of Feynman's ratchet, constructed from the bacterial flagellar motor. The flagellar ratchet makes use of high temperature or water flow to pump protons[9]. **f,** Basic principle of phononics. Three materials A, B and C have different phonon mode density vs frequency curves. Thermal conductance across their interfaces ($I_{AB}$ and $I_{BC}$) increases with curve overlap. Strong temperature dependence of the curve of material B makes a thermal diode possible, which conducts heat in one direction. If C is warmer than A, numerous heat-carrying phonons cross $I_{BC}$ and $I_{BA}$, whereas if A is warmer than C, much less heat-carrying phonons can transport heat in the opposite direction. More complicated systems that are analogs to electronic semi-conductors have been constructed[5].

The discipline of non-equilibrium thermodynamics deals with phenomena that require a thermal gradient. Only recently have practical application of such phenomena started being investigated in solid state physics[5] and the modelling of evolution as in thermal gradients near SHV[11] (Fig. 1a-b). In this study I model the emergence of major organisms in the proposed evolutionary nursery of the so-called 'Garden of Prandtl' (Fig. 1c). Analogous to Feynman's ratchet[10] (Fig. 1d), the bacterial flagellar ratchet (Fig. 1e) would have worked not only on a thermal gradient, but also on flowing water, which can mimic a high temperature. The MT (Fig. 2a) of the eukaryotic cell[12] would be analogs of phononic devices[5] (Fig. 1f).

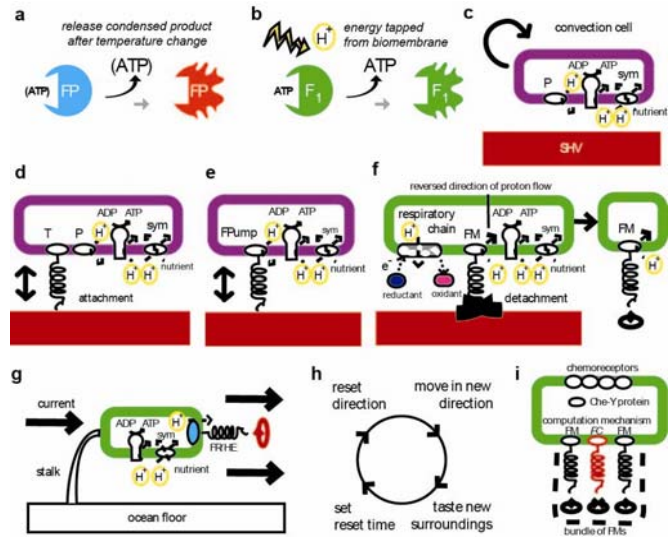

**Figure 2. Conjectured emergence of the bacteria[9]. a**, The proposed First Protein (FP)[3] condensed substrates during a thermal cycle, binding for instance ADP and phosphate at one temperature in its dehydrating cavity, and releasing the formed condensed high-energy product ATP at another, higher (hot denaturation) or lower temperature (cold denaturation). Other possible products '(ATP)' are phospholipids, peptide bonds, and phosphodiester bonds. **b**, Today's ATP Synthase[3] isothermally releases similarly synthesized and bound ATP during high-energy proton transfer across a biomembrane. **c**, In a convection cell, a primordial proton pump P driven by thermal cycling sustained a proton gradient that drove ATP synthesis and nutrient uptake by $H^+$/nutrient symport (sym)[9]. **d,** A filamentous protein attached a bacterium to a SHV and evolved into a thermotether T that oscillated by contraction by cold denaturation. **e**, Merger of thermotether and proton pump in a flagellar pump (Fpump)[9]. **f**, Upon acquisition of proton pumping respiration, the flagellar pump turned into the proton-driven bacterial flagellar motor (FM) and disconnected from the SHV. **g,** The flagellar ratchet/heat engine (FR/HE) pumped protons on high temperature or water flow analogous to Feynman's ratchet (Fig. 1d). It emerged after attachment by a stalk to the isothermal ocean floor or to a SHV (not shown). **h**, The bacterial chemotaxis cycle by which a bacterium moves towards food. (1) By a bundle of flagella, the bacterium moves in a certain direction; (2) It senses whether the concentration of food increases; (3) Slowing or speeding up an internal clock, the direction reset is delayed; (4) Upon a reset, the bacterium tumbles and moves in another direction. **i,** The reset: the bundle of flagellar motors (FM) that surrounds the flagellar computer (FC) forces it to rotate. The FC momentarily functions as brake of the entire bundle at the end of the delay.

Previous studies[3][13] implicitly stated the first conjecture. Therein the emergence was modeled of the genetic machinery including the genetic code, photosynthesis and respiration[3] and the bacterial flagellar motor[9] (Fig. 2c-2f). Life started with a First Protein (FP) (Fig. 2a), a progenitor of the β-subunit of the $F_1$ moiety of ATP Synthase (Fig. 2b) able to perform several condensation reactions during thermal cycling (estimated required minimal temperature amplitude, ~7,5 C[13]). A modified flagellar motor functioned analogous to Feynman's ratchet[9] (Fig. 2g). Prokaryotes possess intelligence: during chemotaxis they act upon gathered information, detecting and moving towards food[9]. A food concentration increasing during movement delays the reset that causes the bacterium to tumble and to randomly change direction (Fig. 2h-2i). Application of the evolutionary continuity principle to the transition from prokaryotic to eukaryotic flagellum or 'cilium' leads to the postulate that this cilium has also been able to gain energy (as ATP) from flow and is involved in computation[9].

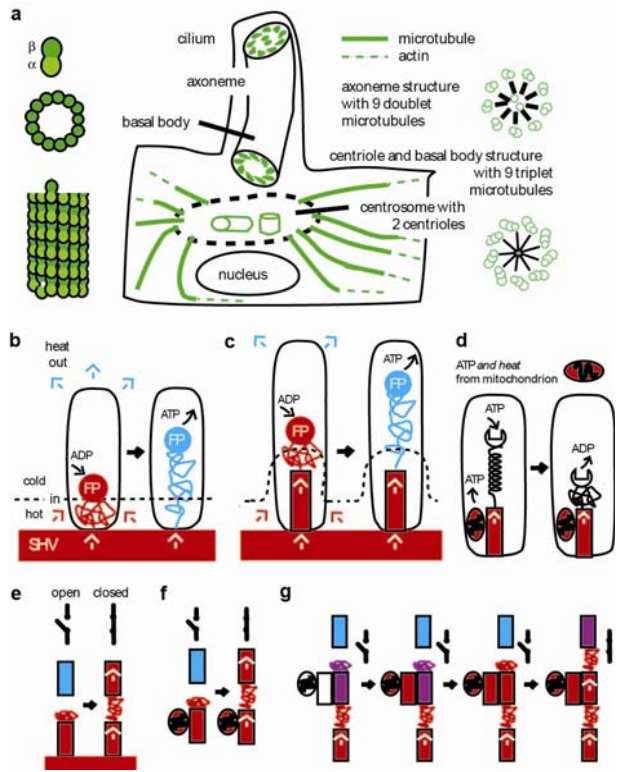

**Figure 3, Conjectured emergence of the eukaryotic cell. a,** The cytoskeleton of the eukaryotic cell contains tube-shaped MT that connect to actin and traverse the cell[12]. The word 'cytoskeleton' is somewhat misleading, as it suggests that the functions of the MT have been identified and are static, while their functions have not yet been completely elucidated and MT are highly dynamic. The centrosome[12] is the center of the MT network and a plausible candidate site of intracellular data processing, also because it is situated between the cilium and the nucleus. Its centrioles resemble the intercellular end ('basal body') of the cilium and contain triplet MT, the cilium doublet MT. **b,** During emergence the thermotether was present inside the cell where it functioned as progenitor of actin. The internal thermotether-FP combination generated ATP in the internal thermal gradient[3]. **c,** The added MT supplied heat to the thermotether/actin. **d**, Heat-generating mitochondria found today adjacent to MT[14] supply heat to them. **e**, Thermal contact between two MT effected by actin functioning as thermal switch. As information processing in man-made computers is based on switches, complex thermal switches based on MT/actin/myosin may have been the fundament of intracellular data processing. **f**, Thermal switch driven by heat generated by a mitochondrion. **g**, Amplification of a temperature signal by a switched-on mitochondrion.

Cavalier-Smith has associated the origin of the eukaryotic cell with the emergence of MT[15]. Today's eukaryotic cell (Fig. 3a) makes use of the mitochondrion —an obvious early endosymbiont—for power (ATP) generation by respiration. The power source of its ancestral host, the 'Archezoan', was according to the second conjecture a thermal gradient in which progenitors of the cytoskeleton (MT/actin) conducted heat to an FP oscillating inside the cell (Fig. 3b-3c). A role of heat transfer in the eukaryotic cell is suggested in analogy to the emerging discipline of phononics[5] and also by: (1) Observation of intracellular thermal gradient generation[16]; (2) Resemblance of MT to carbon-nanotubes[6] with a very high conductivity of 3500 W/(m K)[7]; (3) MT consist of proteins, which in general have a low conductivity but recently the high value of 416 W/(m K) was reported for spider silk protein[8]. Since few measurements have been done, and many proteins exist, many more proteins with a high conductivity may exist; (4) Mitochondria producing heat during thermogenesis are often situated adjacent to MT (Fig. 3d)[14]; this heat could amplify[17] temperature signals (Fig. 3g). Amplification makes thermal switches possible (Fig. 3e-g), which similar to electronic relays in electronic circuits could be applied in thermal circuits that process data by switching (Boolean) algebra. Different protein conformations will have different conductivities, permitting easily read-out information storage. Biological computation could therefore invoke information representation by temperature and conductivity. The heat shock response ubiquitous in eukaryotes could be an adaptation to protect such thermal computation against external temperature disturbances.

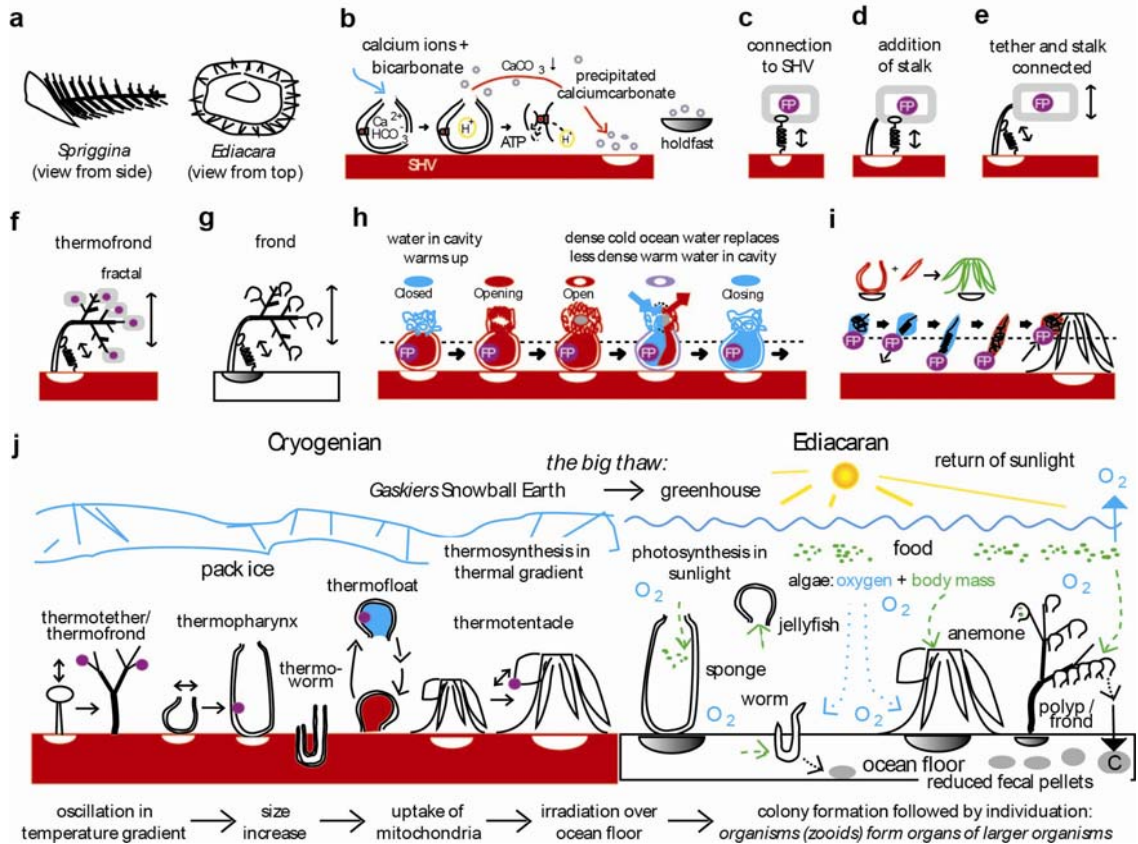

**Figure 4. Conjectured emergence of the metazoans. a,** Cm-sized fossils of typical first animals: *Spriggina* is an extinct frond, *Ediacara* an anemone.[18] **b,** The *Twitya* disks[18] are attributed to organisms that used temperature-stimulated $CaCO_3$ precipitation by the reaction $Ca^{2+} + HCO_3^- \rightarrow CaCO_3 \downarrow + H^+$ to produce high-energy protons (with phosphate a similar process is possible). The precipitate functioned as holdfast. **c,** Attachment of thermotether to holdfast. The purple-coloured FP was thermally cycled. **d,** Attachment of stalk. **e,** Connection of thermotether to stalk yielded a small engine working on the thermal gradient that swept the main body through the thermal gradient. **f,** Stalk size increase by a fractal structure yielded the fan-like thermofrond with attached symbionts. **g,** Adaptation to an environment without a thermal gradient yielded the Ediacaran frond. **h,** The thermopharynx consisted of a sac that at one end connected to the holdfast; on the other end, it contained a toroid-shaped throat (pharynx) that closed by cold denaturation when cold but opened when warm. Subsequent warming of the ingested water by heat from the SHV opened the throat. External dense cold water replaced the less dense warmed water by gravity, and the cold reclosed the throat. The internal lining of the sac contained FPs that generated ATP during the temperature oscillation. **i,** Thermotentacle: Attachment to the thermopharynx of a tentacle that contracted at high temperature (hot protein denaturation) when it approached the SHV yielded another thermal oscillator, as the tentacle contracted when it approached the vent. **j,** Overview. Thermoworm and thermofloat functioned similar to the thermopharynx. In the Garden of Prandtl, size gave a selective advantage, permitting thermosynthesis at a larger distance from the edge. At the end of the Snowball Earth the Cryogenians started to feed on the returned photosynthesizing algae, yielding the Ediacaran metazoans.

The fossils of the first macroscopic animals date from the Ediacaran (650-543 Mya) (Fig. 4a), right after the Gaskiers Snowball Earth, a global glaciation during which photosynthesis was globally impeded[18]. According to the third conjecture, the Cryogenian ancestors of the Ediacarans lived on the thermal gradient between SHV and the cold ocean, using four thermal physiologies: (1) calcification ($CaCO_3$ precipitation) at high temperature with the resulting protons[19] generating ATP by chemiosmosis (Fig. 4b); (2) oscillation in the gradient directly above the SHV by the emerging thermofrond (Fig. 4c-f), ancestor of the Ediacaran fronds (Fig. 4g); (3) heating of ingested cold water by the thermopharynx, ancestor of the sponges (Fig. 4h); and (4) oscillation high-up in the gradient by suspended collagen-containing tentacles (Fig. 4i) part of the thermotentacle, ancestor of the anemones. In small organisms the low Reynolds number hampers movement, the 'Purcell barrier'[9]. Movement sustained by the thermal gradient of the SHV broke the barrier, whereafter size increased, as size frequently gives a selective advantage. In protists, symbiosis, colony formation and individuation are common[3]; the holdfast remained a part of following organisms, and addition of an oscillating tentacle to the thermopharynx yielded the thermotentacle.

Consistent with the notion that the organs of the first animals had a thermosynthesis ancestry, the first Ediacarans lived near deep-water volcanoes[18]. Upon leaving the SHV they kept the acquired capability to transport large volumes of water, reactivated respiration by mitochondria and preyed on smaller organisms such as reemerged algae in the water around them instead of living on temperature differences in the water. Sponges, anemones, worms and jellyfish appeared (Fig. 4j). The large size of their fecal pellets hampered complete decomposition by bacteria, which led to the storage of reduced biomass in the ocean floor with concomitant oxygenation of the atmosphere[20]. This oxygenation made in turn larger animals possible. In the Cambrian (543 Mya) still larger animals emerged, due to the emerged nerve, which enabled signal transfer over a larger distance between the different transport systems constituted by symbionts and thus larger colonies. The origin of the nerve was also thermal: the thermonerve progenitor worked similar to the thermogalvanic cell[3]. Increasing size and oxygenation triggered the Cambrian explosion with its massive self-modifying global changes in geology and biology.

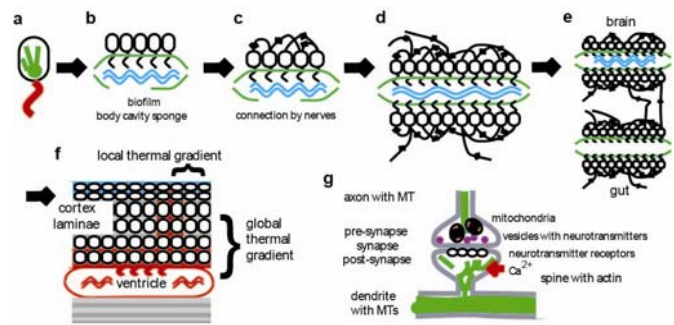

**Figure 5**. **Conjectured emergence of the mammalian brain.** The previous existence of eukaryotic flagellar ratchets is postulated[9] in which water flow was equivalent to a high temperature, just as in Feynman's ratchet (Fig. 1d). The amino acids detected during bacterial chemotaxis later functioned as neurotransmitters. **a**, During chemotaxis the eukaryotic cell functioned similar to the prokarotic cell (Fig. 2h-i). Use of a reset was conserved during the evolution of thermotaxis[21] and during overall functioning[9]. **b**, The biofilm on the Proterozoic ocean gained free energy from the water flow above it[9]. **c**, Biofilms acquired nerves (lines with arrows). **d**, Connection of the entire organism (sponge) by the nerve-containing network around the gut: a central nervous system emerged. **e**, Split in (1) enteric nervous system[22] and (2) brain that collected and processed external data. **f**, Addition of cell layers to the brain: possibly by regression to ancient thermal mechanisms, use was made of thermal gradients between the ventricles and the cortex and inside the cortex. **g**, Today's synapse may still sustain an extracellular thermal switch. Temperature related structures are: mitochondria that can produce heat by thermogenesis, $Ca^{2+}$ channels that can sustain a $Ca^{2+}$ influx that increases the temperature[16], microtubule/actin system in the dendrite/spine which resembles the MT/actin combination shown in Fig. 3c, and spines increasing or decreasing (even vanishing) in strength concomitantly with the temperature[23].

The underlying thermal machinery of the brain proposed in the forth conjecture invokes (1) intracellular thermal switches (Fig. 3g-h), (2) extracellular thermal switches (Fig. 5g), (3) thermal effects on the cilium, and (4) the thermal diffusion potential as in the thermonerve. The following list of observations supports the conjecture: (1) brain death correlates with absence of a global thermal gradient between the inside and outside of the brain[24]; (2) consciousness requires the global gradient, which decreases during sleep; (3) brain cooling is applied as anesthetic during brain surgery; (4) local cortex activation is associated with a local temperature increase[25] and a bird's song can be switched off by local brain cooling[26]; (5) diffusion MRI detects a high temperature in the ventricles[27], the ependyma lines the ventricles, and the ependymal cilia[28] span the thermal gradient; (6) in embryoes, fluid flow over the ependyma induces electrical activity in the cortex[29]; (7) on a smaller scale, actin is a major component of the dendritic spines of neurons. Spine plasticity is important for learning and memory[30]. A recent review of thermal effects mentions observed loss of spines at low, and spine potentiation at high temperature[23], consistent with its functioning as a thermal switch affected by thermal gradients. There is consequently much evidence for a functional role of thermal gradients in the brain.

Aristotle systematically invoked heat in his work and associated in *On the Motion of Animals* thought with local temperature changes. Physicists and engineers nowadays apply heat widely, both experimentally and theoretically. This study challenges the implicit 'isothermal life paradigm' that life and evolution do not or did not invoke heat in major processes—a tenet that life scientists have internalized *but that is unwarranted*. They consider the numerous biological phenomena involving thermal cycling or a thermal gradient as accidental, unrelated and not fundamental: heat shock, thermogenesis, the H2A.Z-histone, the chromatin thermostat, CG islands during DNA melting, circadian rhythms, organism presence in thermal gradients (rock:air, soil:air, snow:air, sea:air, sea:ice, volcanic rock:water, ocean fronts, Langmuir circulations, thermoclines, SHV), thermotaxis (of bacteria, of nematodes, of sharks), torpor, hibernation, the radiator hypothesis, the hemo-neural hypothesis. Many thermal phenomena relate to propagation: synchronized division, phenology, vernalization, germination, flowering, flower ovens, budding, oviduct, ovulation, estrus, scrotum, orgasm, fertilization, moulting, diapause breaking. Haeckel's biogenetic law may apply to these phenomena, with early stages of ontology recapitulating the first stage of phylogeny, the origin of life.

The general conjecture turns established ideas in biology upside down: instead of considering heat as a product of secondary importance, essentially waste, it gives primacy to heat's general initiating role usable for power generation and signalling. Where proposed related processes are extant, *in situ* experiments could test the conjecture—where extinct, the methods of synthetic biology could be applied to recreate the conjectured lost evolutionary scaffolds. It seems particularly promising to investigate (1) thermal gradients inside the eukaryotic cell[16], (2) the thermal conductivity of today's MT, (3) thermotaxis of *C. elegans*[21], and the proposed evolutionary paths that percolate life: (4) *Feynman ratchet → bacterial flagellum → eukaryotic flagellum* (*cilium*) *→ ependymal cilium* and (5) *thermotether → thermal switch → actin → dendritic spine*. It is concluded that the thermal phenomena observed in biology can be combined with the discipline of non-equilibrium thermodynamics in novel and testable 'gestalts' of how life evolved in the past and how it functions today. Life at all times and at all levels may have been entwined with thermal physics.

Acknowledgements
For helpful discussions I thank Steph Menken, Roel van Driel, Paul Fransz, Paul Verbruggen, Dirk Schulze-Makuch, Rudolf Sprik and Hans Westerhoff. The extensive support of the library staff Dyoke van Assum, George Meerburg and Marijke Duyvendak is also recognized.

Author Information
Reprints and permissions information is available at www. nature.com/reprints. The author declares no competing financial interest. Correspondence and requests should be addressed to A.M. (a.w.j.muller @ uva.nl).